\documentclass[12pt,a4paper]{article}
\pdfoutput=1

\usepackage{jcappub}

\usepackage{graphicx}
\usepackage{amsmath}
\usepackage{amssymb}

\def\bea{\begin{eqnarray}}
\def\eea{\end{eqnarray}}
\def\ba{\begin{eqnarray}}
\def\ea{\end{eqnarray}}
\def\be{\begin{equation}}
\def\ee{\end{equation}}
\def\beq{\begin{equation}}
\def\eeq{\end{equation}}
 
\newcount\hour \newcount\minute
\hour=\time \divide \hour by 60
\minute=\time
\title{Halo-independent tests of dark matter direct detection signals: local DM density, LHC, and thermal freeze-out}

\def\kth{Department of Theoretical Physics,
School of Engineering Sciences, KTH Royal Institute of Technology,
AlbaNova University Center, 106 91 Stockholm, Sweden}

\def\okc{Oskar Klein Centre for Cosmoparticle Physics, Department of Physics,
Stockholm University, SE-10691 Stockholm, Sweden}

\author[a]{\textbf{Mattias Blennow,}\vspace*{0mm}}

\author[a]{\textbf{Juan Herrero-Garcia,}\vspace*{0mm}}
\affiliation[a]{\kth}

\author[b]{\textbf{Thomas Schwetz}\vspace*{0mm}}

\author[b]{\textbf{and Stefan Vogl}\vspace*{0mm}}
\affiliation[b]{\okc}

\emailAdd{emb@kth.se}
\emailAdd{juhg@kth.se}
\emailAdd{schwetz@fysik.su.se}
\emailAdd{stefan.vogl@fysik.su.se}

\abstract{
From an assumed signal in a Dark Matter (DM) direct detection
experiment a lower bound on the product of the DM--nucleon scattering
cross section and the local DM density is derived, which is
independent of the local DM velocity distribution. This can
be combined with astrophysical determinations of the local DM density. Within a given particle physics model the bound also allows a robust comparison of a
direct detection signal with limits from the LHC. Furthermore, the bound can be used to formulate a condition which has to be fulfilled if the particle responsible for the direct detection signal is a thermal relic, regardless of whether it
constitutes all DM or only part of it. We illustrate the arguments
by adopting a simplified DM model with a $Z'$ mediator and assuming a
signal in a future xenon direct detection experiment.}


\begin{document}
\maketitle

\flushbottom
%%%%%%%%%%%%%%%%%%%%%%%
\section{Introduction} 
%%%%%%%%%%%%%%%%%%%%%%%

We know from gravitational effects that dark matter (DM) constitutes a
significant fraction of the energy density in the universe. One of the
most promising ways to directly detect it is to look for the
scattering of DM particles from the galactic halo in underground
detectors \cite{Goodman:1984dc, Drukier:1986tm, Freese:1987wu}. In the interpretation of these
direct detection (DD) signals, the astrophysical input plays a crucial
role. Typically, the velocity distribution of DM is assumed to be a
Maxwellian distribution truncated at the galactic escape velocity,
$v_{\rm esc}$, known as the Standard Halo Model (SHM). For a given
halo model and a particle physics model for the DM--nucleus
interaction, a positive direct detection signal will provide an
allowed region in the dark matter mass ($m_\chi$) vs.\ cross section
plane ($\sigma_{\rm SI/SD}$). However, using the SHM is very likely an
oversimplification, with $N$-body simulations indicating a more
complicated structure of the DM halo, see for instance
refs.~\cite{Vogelsberger:2008qb, Kuhlen:2009vh, Kuhlen:2012fz}.

Therefore, in order to interpret DD signals, halo model independent methods
have been developed \cite{Drees:2007hr, Drees:2008bv, Fox:2010bz,
  Fox:2010bu, McCabe:2011sr, McCabe:2010zh, Frandsen:2011gi,
  HerreroGarcia:2011aa, HerreroGarcia:2012fu, DelNobile:2013cta,
  DelNobile:2013cva, Bozorgnia:2013hsa, Cherry:2014wia, Fox:2014kua,
  Feldstein:2014gza, Feldstein:2014ufa, Bozorgnia:2014gsa,
  Anderson:2015xaa,Scopel:2014kba,Kavanagh:2012nr,Kavanagh:2013wba}.
Most of these use the fact that for a
given particle physics model one can compare the results of different
direct detection experiments without the need of specifying the total
scattering cross section, the local DM density, the galactic escape
velocity, nor the velocity distribution.  In
ref.~\cite{Blennow:2015oea} those methods have been extended to the
comparison of a DD signal and a neutrino signal from DM annihilation
inside the Sun. In the present paper we show how a positive signal
from a DD detection experiment can be used to place a lower bound on
the product of the local DM density $\rho_\chi$ and the scattering
cross section, independent of the DM velocity
distribution. Within a given particle physics model such a lower bound
can be compared to upper limits from LHC as well as to the hypothesis
of DM production via thermal freeze-out in the early Universe.

This paper is structured as follows. After setting
the notation for direct detection in sec.~\ref{sec:DD}, we derive 
various inequalities involving the halo integral in
sec.~\ref{sec:eta_bound}. In sec.~\ref{sec:bound} we apply those
bounds to a positive signal in a direct detection experiment, leading
to a lower bound on the product of the local DM density and the
scattering cross section. Those bounds are independent of the DM
velocity distribution, and we discuss various versions of the bound,
highlighting the advantages and disadvantages of the different bounds
in the case of the 3 DM candidate events observed in the CDMS
experiment, as well as for mock data from a signal in a future direct
detection experiment. In secs.~\ref{sec:LHC} and \ref{sec:relic} we
adopt a so-called simplified DM model and show how the bounds from a
DD experiment can be correlated with limits from LHC and with the
thermal freeze-out hypothesis. We conclude in
sec.~\ref{sec:conclusions}.

%%%%%%%%%%%%%%%%%%%%%%%%%%%%%%%%%%%%%%%%%%%
\section{Dark matter direct detection} 
%%%%%%%%%%%%%%%%%%%%%%%%%%%%%%%%%%%%%%%%%%%
\label{sec:DD}

In this section we review the relevant expressions for DD of dark
matter. We focus on elastic scattering of DM
particles $\chi$ with mass $m_\chi$ off a nucleus with mass number $A$
and mass $m_A$, depositing the nuclear recoil energy $E_R$.
The differential rate 
for a 
detector consisting of different target nuclei is given by:
\begin{equation}\label{eq:R}
  \mathcal{R}(E_R, t) = \frac{\rho_\chi}{m_\chi}\,\sum_A \frac{f_A}{m_A}
  \int_{|\vec{v}| > v_m^A} d^3 v  \, v f_{\rm det}(\vec{v}, t) \frac{d \sigma_A}{d E_R}(v)  \,,
\end{equation}
where $\rho_\chi$ is the local DM mass density, $f_A$ corresponds to the mass fraction of nuclei $A$ in the detector, and 
\begin{equation}\label{eq:vm}
v_m^A=\sqrt{\frac{m_A E_{R}}{2 \mu_{\chi A}^2}}  
\end{equation}
is the minimal
velocity of the DM particle required for a recoil energy $E_{R}$, where $\mu_{\chi A}$ is the reduced mass of the DM--nucleus system. For single target detectors, there is just one contribution and thus the sum over $A$ is absent. 
The function $f_{\rm det}(\vec{v}, t)$ describes the distribution of
DM particle velocities in the detector rest frame, with the normalization 
$\int  d^3 v \, f_{\rm det}(\vec{v}, t) =1$. The velocity
distributions in the rest frames of the detector, the Sun and the
galaxy are related by 
$
f_{\rm det}(\vec{v},t) = f_{\rm Sun}(\vec{v} +
\vec{v}_e(t))=f_{\rm gal}(\vec{v} + \vec{v}_s+\vec{v}_e(t)) \,, 
$
where $\vec{v}_e(t)$ is the velocity vector of the Earth relative to
the Sun and $\vec{v}_s$ is the velocity of the Sun relative to the
galactic frame. In the following we are going to ignore the small time
dependence of the event rate due to $\vec{v}_e(t)$ and work in the
approximation of $f_{\rm det}(\vec{v}) \approx f_{\rm gal}(\vec{v} +
\vec{v}_s)$ being constant in time.\footnote{Bounds similar
    to the ones presented below based on the annual modulation signal
    can be found in ref.~\cite{Herrero-Garcia:2015kga}.}

To be specific, in the following we will concentrate on
spin-independent (SI) and spin-dependent (SD) scattering from a
contact interaction. This implies that the differential scattering
cross section $d\sigma_A(v)/dE_R$ scales as $1/v^2$. For SI contact
interactions with equal DM couplings to neutrons and protons the cross
section becomes
\begin{equation}\label{eq:CS}
  \frac{d\sigma_A}{dE_R}(v) = \frac{m_A \sigma_{\rm SI} A^2}{2\mu^2_{\chi p} v^2} F^2_A(E_R) \,,
\end{equation}
where $\sigma_{\rm SI}$ is the total DM--proton scattering cross
section at zero momentum transfer, $\mu_{\chi p}$ is the DM--proton
reduced mass, and $F_{A}(E_R)$ is a nuclear form factor.  For SD
interactions a similar formula applies with a different form factor
and no $A^2$ enhancement, with the zero-momentum DM--proton scattering
cross section denoted by $\sigma_{\rm SD}$.

The event rate can be written as
\begin{align}  
\mathcal{R}(E_R) = \mathcal{C}\,\sum_A f_A \,A^2  F_{A}^2(E_R) \,
\eta (v_m^A) \,,
\label{eq:R0}
\end{align}
where we have defined
\beq\label{eq:eta} 
\eta(v_m^A) \equiv \int_{v>v_m^A}  d^3 v\, \frac{f_{\rm det} (\vec{v})}{v},\qquad \mathcal{C} \equiv  \frac{\rho_\chi \sigma_{\rm SI} }{2 m_\chi\mu_{\chi p}^2}.
\eeq

For a specific detector the number of DM induced events in an energy range 
between $E_1$ and $E_2$ is given by 
\begin{equation} \label{Nevents} 
N_{[E_1,E_2]} = M \,T\, \mathcal{C}\, \langle \eta  (v_m^A) \rangle_{E_1}^{E_2} \,,
\end{equation} 
where $M$ and $T$ are the detector mass and exposure time,
respectively, and we introduce the short-hand notation for energy
integration and target nucleus weighted sum of a quantity $X(v_m^A)$ as
\begin{equation}\label{eq:shorthand}
\langle X \rangle_{E_1}^{E_2}
\equiv 
\sum_A f_A \,A^2
\int_0^\infty d E_R \, F_{A}^2(E_R)\, G_{[E_1, E_2]}^A(E_R) \, X\,,
\end{equation}
where $G_{[E_1,E_2]}^A(E_R)$ is the detector response function
describing the probability that a DM event with true recoil energy
$E_R$ is reconstructed in the energy interval $[E_1,E_2]$, including
energy resolution, energy dependent efficiencies, and possibly also
quenching factors.\footnote{Note that $\langle X \rangle_{E_1}^{E_2}$
  is not an average. We use this notation to indicate energy
  integration and sum over targets.}

%%%%%%%%%%%%%%%%%%%%%%%%%%%%%%%%%%%%%%%%%%%%%%%%%%%%%%%%%%%%%%5
\section{Bounding the halo integral}
%%%%%%%%%%%%%%%%%%%%%%%%%%%%%%%%%%%%%%%%%%%%%%%%%%%%%%%%%%%%%%5
\label{sec:eta_bound}

\begin{figure}
	\centering
	\includegraphics[width=0.65\textwidth]{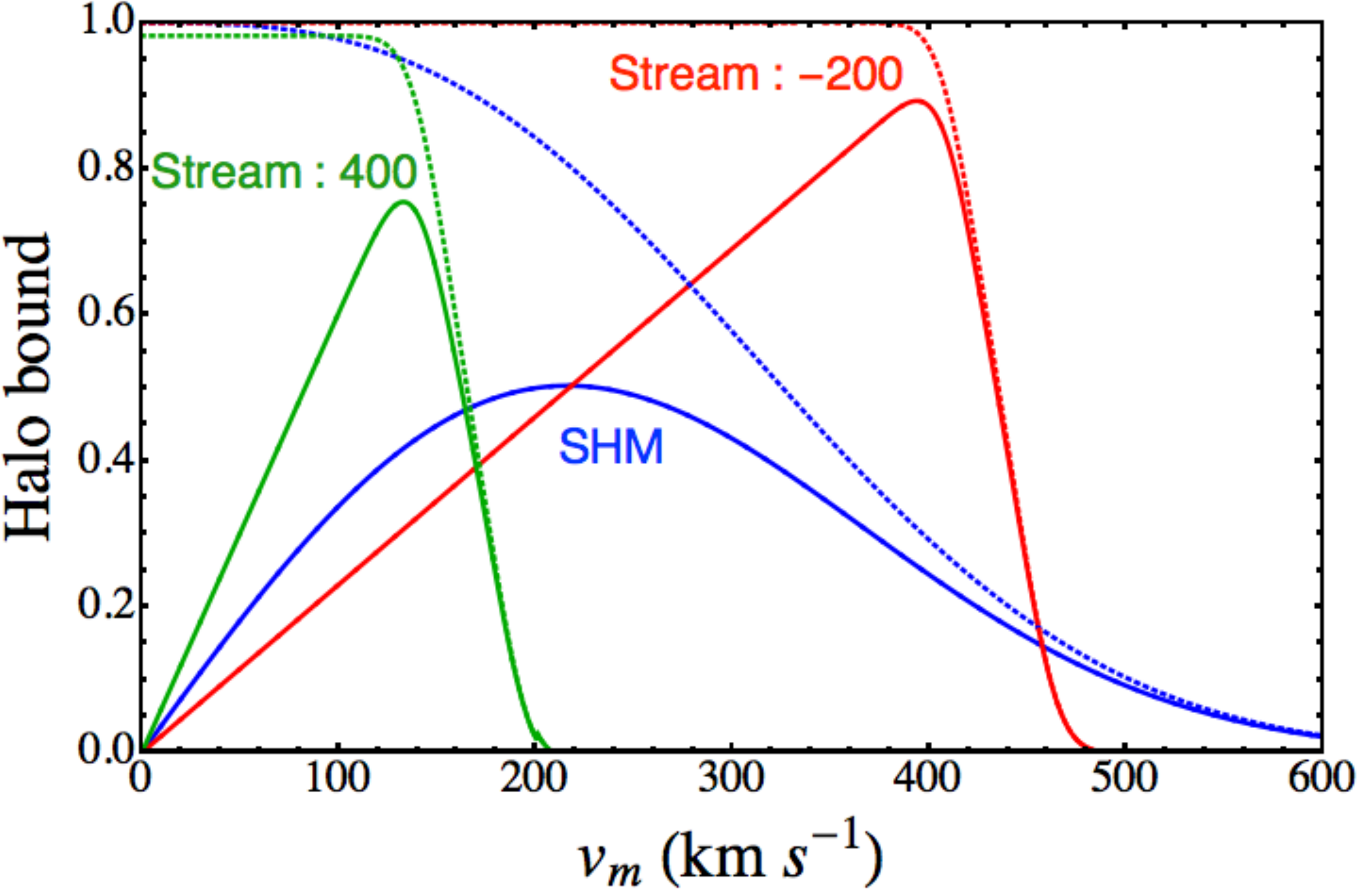}		
	\caption{ Strength of the bound on the halo integral versus
          $v_m$ for the SHM (blue) and for two DM stream examples
          (red, green). We take the streams to be aligned with the
          motion of the Sun in the galaxy, where the velocities of the
          streams relative to the galaxy are chosen to be $-200$ (red)
          and $+ 400$~km/s (green), see labels in the plot. The
          velocity dispersion of the streams is taken to be
          20~km/s. The solid curves show the product $v_m \eta(v_m)$,
          corresponding to the ratio of the left and right-hand sides
          of eq.~\eqref{eq:bound}. The dotted curves correspond to the
          right-hand side of eq.~\eqref{eq:bound_FK} with $v_1 = v_m$
          and $v_2 \to \infty$.} \label{vmeta}
\end{figure}

An upper bound on the halo integral $\eta(v_m)$ defined in
eq.~\eqref{eq:eta} can be derived in the following way (see
  also ref.~\cite{Kavanagh:2012nr}):
\begin{align}  
\eta(v_m^A)& \equiv \int_{v>v_m^A}  d^3 v\, \frac{f_{\rm det} (\vec{v})}{v} \nonumber \\
& \leq \frac{1}{v_m^A}\, \int_{v>v_m^A}  d^3 v\, f_{\rm det} (v) \nonumber \\
&\leq \frac{1}{v_m^A} \label{eq:bound}
\end{align}
where in the last step we used that $\int_{v_m}\,H(v)\,dv\leq \int_0 H(v)\,dv$ for any positive function $H(v)\geq0$, and the normalization condition. While the
inequality is completely general and holds for all possible velocity
distributions it will be useful only if it is not very far from being
saturated, or in other words, if the ratio between the true value of
$\eta(v_m^A)$ and $1/v_m^A$ is not too small. In fig.~\ref{vmeta} we
show with solid curves the product $v_m \eta(v_m)$ for the SHM as well as for two cold
DM stream examples. If this product is close to one the inequality
\eqref{eq:bound} is saturated and if it is much smaller than one the
bound is weak. For the SHM\footnote{\label{shm}Here and in the following we adopt
  the following parameters for the SHM: we use a Maxwellian velocity
  distribution with the mean velocity $\bar v = 220$~km/s, truncated
  at the escape velocity of $v_\mathrm{esc} = 550$~km/s.} 
  one can see that the lower bound is reasonably strong in the $v_m$ range
between $50$ and $500\,{\rm km \,s^{-1}}$ and gets weak for low and
high $v_m$ values. An important point one should keep in mind is that,
at very low DM masses, the $v_m$ values relevant for DD can be much
larger that the expected escape velocities in the detector rest-frame,
$\sim750\, \rm km\,s^{-1}$ (there are large uncertainties, see for
instance refs.~\cite{McCabe:2010zh, Frandsen:2011gi,Lavalle:2014rsa}). The
upper bound in eq.~\eqref{eq:bound} is going to become weak in
scenarios where the high-velocity tail is probed, e.g.\ for low dark
matter masses or high thresholds.

For streams the bound gets strong for $v_m$ close to the velocity of
the stream in the detector frame. This is obvious from
eq.~\eqref{eq:bound} by approximating a DM stream by $f(\vec{v})
\propto \delta^3(\vec{v} - \vec{v}_{\rm stream})$. In this
approximation $v_m \eta(v_m)$ is a linear function rising up to
1 at $v_m = v_{\rm stream}$. This behavior is visible in
fig.~\ref{vmeta}. The curves do not rise up to 1 because of the
finite velocity dispersion of 20~km/s assumed in the calculations.

In ref.~\cite{Feldstein:2014ufa} a lower bound on the DM cross section
was derived based on a different inequality for the halo
integral. Having DD experiments sensitive to the
velocity range between $v_1$ and $v_2$ in mind, one obtains the inequality~\cite{Feldstein:2014ufa}
\begin{equation}
  1 = \int_0^\infty dv \, \eta(v) 
    \ge v_1 \eta(v_1) + \int_{v_1}^{v_2} dv \, \eta(v) \,,
\label{eq:bound_FK}
\end{equation}
where the first identity follows from the normalization of $f_{\rm
  det}(\vec{v})$ and the inequality follows from the fact that
$\eta(v)$ is a non-negative monotonously decreasing function of $v$.
We observe that eq.~\eqref{eq:bound} simply corresponds to the first
term in eq.~\eqref{eq:bound_FK}. The dotted curves in fig.~\ref{vmeta}
show the right-hand side of eq.~\eqref{eq:bound_FK}. We find that, for
high velocities, the two bounds become similar, whereas for low
velocities the inequality in eq.~\eqref{eq:bound_FK} is close to
saturated and is expected to provide stronger bounds than
eq.~\eqref{eq:bound}. Below we will comment on the
advantages/disadvantages of the two bounds when applied to data.

Velocity distributions obtained from N-body simulations are
  qualitatively similar to the SHM, although quantitative differences
  occur, see e.g.~\cite{Vogelsberger:2008qb, Kuhlen:2009vh,
    Kuhlen:2012fz}. Hence, the strength of the bounds for such
  velocity distributions is expected to be similar to the SHM case
  shown in fig.~\ref{vmeta}. Note also that a hypothetical dark matter
  disk effectively corresponds to a DM stream, and thus we expect also
  qualitatively a similar behaviour as for the streams shown in the
  figure.

%%%%%%%%%%%%%%%%%%%%%%%%%%%%%%%%%%%%%%%%%%%%%%%%%%%%%%%%%%%%%%5
\section{A velocity-distribution-independent lower bound on 
$\rho_\chi \sigma_{\rm SI/SD}$ from a direct detection signal} 
%%%%%%%%%%%%%%%%%%%%%%%%%%%%%%%%%%%%%%%%%%%%%%%%%%%%%%%%%%%%%%%%%%%%%%%%
\label{sec:bound}

We now use the bounds on the halo integral to derive lower bounds on
the product of DM density multiplied by the scattering cross section $\rho_\chi \sigma_{\rm SI/SD}$. In
sec.~\ref{sec:CS_our} we will use eq.~\eqref{eq:bound} to derive a bound
based on the number of observed events in a DD experiment, whereas in
sec.~\ref{sec:CS_FK} we will comment on a bound based on
eq.~\eqref{eq:bound_FK}, which is useful if the recoil energy
spectrum of DM scattering events can be measured with high precision. In this section we concentrate on a signal from just one direct detection experiment, but we comment on the multi-experiment case in the conclusions, sec.~\ref{sec:conclusions}.

\subsection{Lower bound from the number of observed events}
\label{sec:CS_our}

Let us now apply the bound eq.~\eqref{eq:bound} to the event rate in a
DD experiment. For definiteness we focus on SI
interactions. The generalization to the SD case is straight-forward.
Inserting the bound from eq.~\eqref{eq:bound} into eq.~\eqref{eq:R0} we obtain
\begin{equation} \label{eq:RAta}
\mathcal{R}(E_R) \leq \mathcal{C} \,\sum_A\,\frac{f_A A^2 \, F_A^2(E_R)}{v_m^A(E_R)} \,.
\end{equation}
With the definition of $\mathcal{C}$ in eq.~\eqref{eq:eta} this may be re-written as a lower bound on 
$\rho_\chi \sigma_{\rm SI}$ which does not depend on $f(v)$:
\beq \label{eq:RAtb} 
\rho_\chi \sigma_{\rm SI} \geq \frac{2 \,m_\chi\,\mu_{\chi p}^2}{\sum_A\,f_A\,A^2\, F_A^2(E_R)/v_m^A(E_R)}\,\mathcal{R}(E_R) \,.
\eeq
This inequality must be fulfilled at all energies $E_R$.
Taking the more realistic situation of a finite energy resolution and other detector effects into account we can also derive a corresponding bound in terms of the measured number of events within an energy interval $[E_1,E_2]$ by use of eq.~\eqref{Nevents}:
\beq \label{eq:RAtd} 
\rho_\chi \sigma_{\rm SI} \geq 
\frac{2 \,m_\chi\,\mu_{\chi p}^2}{MT\,\langle 1/v_m^A\rangle_{E_1}^{E_2} } 
\, N_{[E_1, E_2]} \,,
\eeq
where $\langle 1/v_m^A\rangle_{E_1}^{E_2}$ is defined in
eq.~\eqref{eq:shorthand}.  If a DD experiment reports a lower bound $B_{\rm CL}$ at
some confidence level (CL) on DM induced events in a certain energy
interval, $N_{[E_1,E_2]} > B_{\rm CL}$, then eq.~\eqref{eq:RAtd} provides a lower bound on the
product $\rho_\chi \sigma_{\rm SI}$ at that CL, which is independent of the local DM
velocity distribution.
\bigskip

In the following we use the putative signal from the CDMS silicon
exposure~\cite{Agnese:2013rvf} to illustrate how this bound can be
used.  The CDMS collaboration reports 3 candidate events from their
data with a silicon target, 
rejecting the known-background-only hypothesis with a $p$-value of 0.19\% when tested against the DM+background hypothesis using a profile likelihood ratio test.
Although a DM interpretation of this signal
is in tension with limits from other experiments~\cite{Ahmed:2011gh,
  Aprile:2012nq, Akerib:2013tjd, Angloher:2014myn, Agnese:2014aze, Xiao:2015psa}
(see for instance refs.~\cite{Feldstein:2014ufa, Bozorgnia:2014gsa}
for halo-independent analyses) we use this signal as a case study and
apply eq.~\eqref{eq:RAtd} to it. We use the Helm
parameterization for the SI form factor, $F(E_R) = 3 e^{-q^2
s^2/2} [\sin(q r)-q r\cos(q r)] / (q r)^3$, with $q^2 = 2 m_A E_{R}$, $s
= 1$~fm, $r = \sqrt{R^2 - 5 s^2}$ and $R = 1.2 A^{1/3}$~fm.

\begin{figure}
	\centering
	 \includegraphics[width=0.65\textwidth]{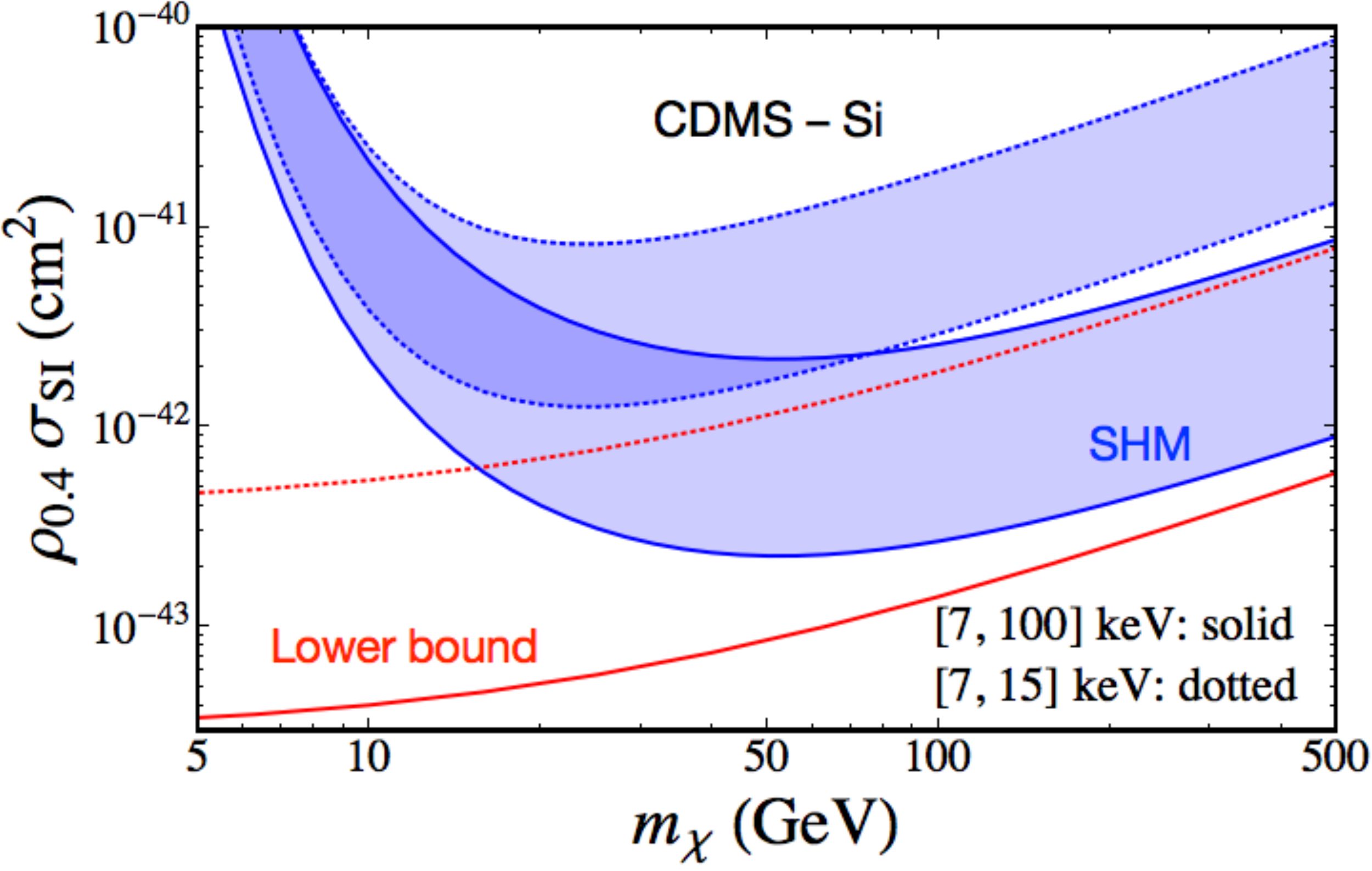}
	\caption{Lower bound (SHM interval) in red (blue) for
          $\rho_{0.4} \, \sigma_{\rm SI}$ from CDMS-Si data versus DM
          mass, where $\rho_{0.4} \equiv \rho_\chi/(0.4\, \rm GeV \,
          cm^{-3})$.  We show the $90\%$~CL results for two different
          choices of the energy range: $[7,15]$ keV (dotted) and
          $[7,100]$ keV (solid).}\label{sigma_bounds}
\end{figure}

The red curves in fig.~\ref{sigma_bounds} show the 90\%~CL lower bound on
$\sigma_{\rm SI}$ from CDMS-Si data for a reference value of
$\rho_\chi=0.4\,\rm GeV/cm^3$. Those can be compared to the
allowed interval for $\sigma_{\rm SI}$ when the SHM is assumed (shown
as blue shaded bands in the plot).\footnote{Note that the SHM region
  is based only on the observed number of events, without using any
  energy information. Therefore, we obtain a degenerate band in
  $m_\chi$, opposed to the closed regions resulting e.g., from an
  event-based likelihood analysis.} The behavior follows from the
discussion related to fig.~\ref{vmeta}. For low DM masses only large
values of $v_m$ are probed by the experiment and the bound becomes
much weaker compared to the SHM, where no DM particles are left with
such high velocities due to the escape velocity cut-off. For DM masses
$m_\chi \gtrsim 100$~GeV the lower bound is close to the SHM
interval. However, we note that for DM masses in that range the
CDMS-Si signal is highly disfavoured by other experiments.

In general, for a given observed event distribution it is not a
priori clear which energy interval will give the strongest constraint, as
the expected spectrum $R_A(E_R)$ decreases with energy, while
$v_m(E_R)$ increases. The form factor typically decreases, but can
also show local minima. This effect is shown for CDMS-Si data in
fig.~\ref{sigma_bounds}, where the results are shown for two different
energy intervals, $[7,15]$~keV and $[7,100]$~keV.  We use the
expected background spectrum from ref.~\cite{McCarthy}. We observe that the smaller energy
interval, $[7,15]$~keV, provides the strongest bound, since in this
case the signal to background ratio is highest.

\begin{figure}
	\centering
  \includegraphics[width=0.65\textwidth]{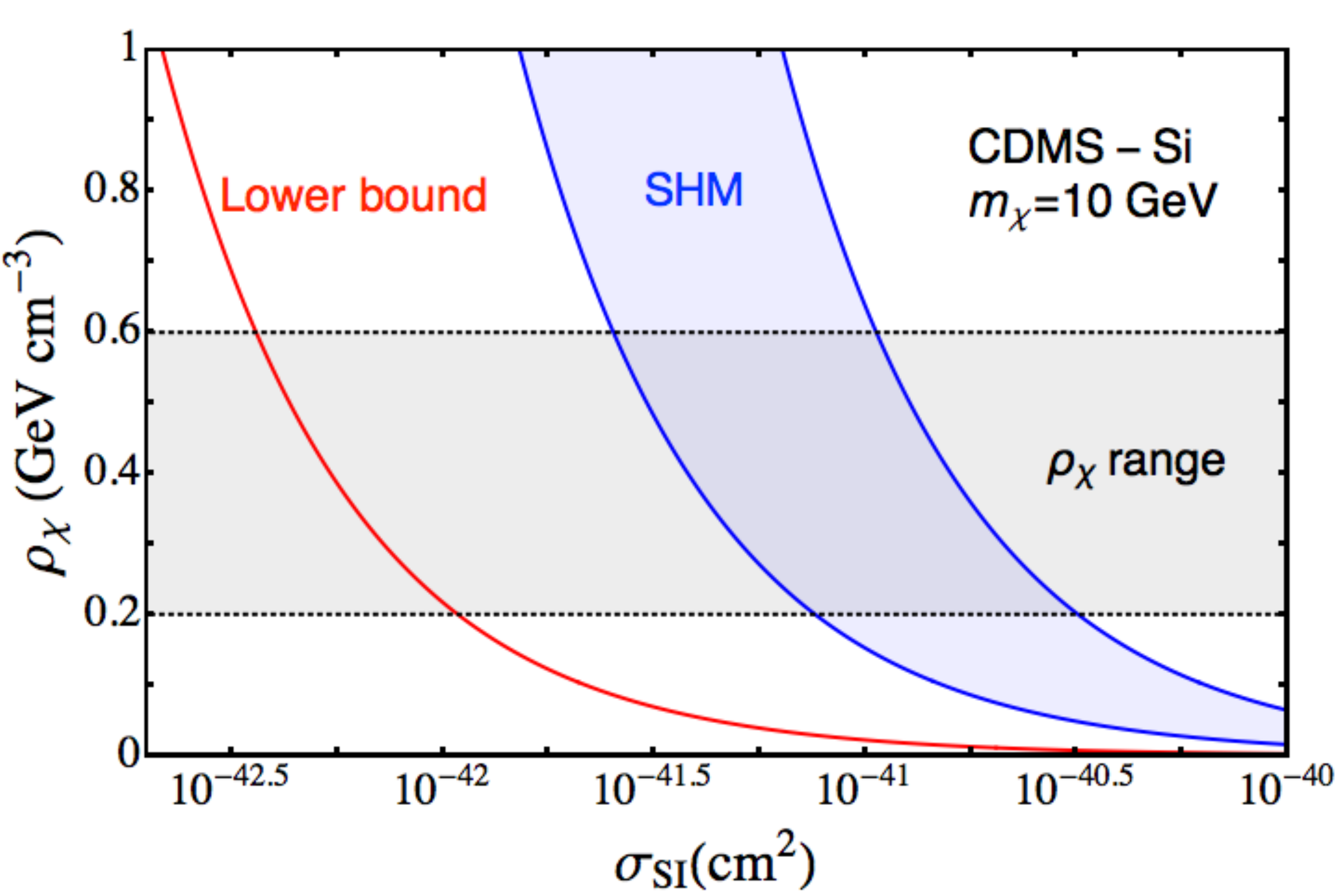}		
  \caption{Lower bound at $90\%$ CL on the local DM density
    $\rho_\chi$ from CDMS data (in red), shown for
    $m_\chi=10$~GeV. The blue shaded region shows the allowed range at
    $90\%$~CL assuming the SHM. The grey shaded horizontal band
    indicates the preferred range for $\rho_\chi$ from Milky Way
    observations.}
    \label{rhoCDMS}
\end{figure}

Another way to use eq.~\eqref{eq:RAtd} is to consider it as a lower
bound on the local DM density $\rho_\chi$ for a given scattering cross
section and DM mass. This lower bound can then be compared to
astrophysical determinations of $\rho_\chi$ to identify regions in
$\sigma_{\rm SI/SD}$ and $m_\chi$ which are compatible with reasonable
values of $\rho_\chi$.  In fig.~\ref{rhoCDMS}, we show for
  illustration the 90\%~CL lower bound on the DM density from CDMS-Si
  data as a function of $\sigma_{\rm SI}$ for $m_\chi=10$ GeV and
  compare it with the 90\%~CL interval obtained from assuming the
  SHM. We use the recoil energy interval of $[7,15]$~keV.
The value for the DM chosen in the figure is motivated by
  the fact that typically for masses in this range the tension of the
  CDMS-Si signal with bounds from other experiments is less
  severe. Corresponding results for different dark matter masses can
  be obtained by recasting the limit on $\rho_{0.4} \, \sigma_{\rm
    SI}$ shown in fig.~\ref{sigma_bounds} into the ($\sigma_{\rm SI},
  \rho_\chi$) plane.

These results can be compared to astrophysical determinations of the
local DM density. There are various methods to infer $\rho_\chi$,
either based on local dynamical tracers \cite{Salucci:2010qr,
  Garbari:2012ff, Zhang:2012rsb} or global methods based on fitting a
mass model of the Milky Way to observations \cite{Catena:2009mf,
  Weber:2009pt, McMillan:2011wd, Iocco:2011jz, Pato:2015dua} (see 
ref.~\cite{Read:2014qva} for a recent review).  Depending on the
different assumptions, values for $\rho_\chi$
roughly in the range between 0.2 and 0.6~GeV/cm$^3$ are found, mostly consistent
within the quoted error bars (with the size of the errors also
strongly dependent on assumptions), see table~4 of ref.~\cite{Read:2014qva}
for a summary. The gray shaded horizontal band in fig.~\ref{rhoCDMS}
indicates the plausible range for $\rho_\chi$, motivated by the studies
quoted above. From the red curve in the figure we observe that cross
sections of $\sigma_{\rm SD} \lesssim 3\times 10^{-43}\, \rm cm^2$ are
disfavoured, since the local DM would
need to be too high to obtain the observed signal for such small cross sections. Note that this
argument also applies to the case when the species $\chi$ constitutes
only part of the DM, since this would only increase the lower bound on
the total DM density.

\subsection{Lower bound from a precise recoil energy spectrum measurement}
\label{sec:CS_FK}

Let us now discuss a bound based on eq.~\eqref{eq:bound_FK}. For a single target experiment with perfect energy resolution, a measurement of the spectrum $\mathcal{R}(E_R)$ allows a determination of the halo integral via eq.~\eqref{eq:R0}:
\begin{align}\label{eq:eta_obs}
\eta (v_m^A) =
\frac{\mathcal{R}(E_R)}{\mathcal{C}\,A^2  F_{A}^2(E_R)} \,.
\end{align}
Consider a spectral measurement of $\mathcal{R}(E_R)$ in the energy
range $[E_1,E_2]$, which for a given DM mass can be related to a
velocity interval $[v_1, v_2]$ via eq.~\eqref{eq:vm}. Inserting
eq.~\eqref{eq:eta_obs} into the bound eq.~\eqref{eq:bound_FK} and
using the definition of $\mathcal{C}$ leads to the lower bound \cite{Feldstein:2014ufa}
\begin{align}\label{eq:boundCS_FK}
  \rho_\chi \sigma_{\rm SI} \ge \frac{2 m_\chi \mu^2_{\chi p}}{A^2} 
  \left( v_1 \frac{\mathcal{R}(E_1)}{F_{A}^2(E_1)} + 
         \int_{v_1}^{v_2}dv \frac{\mathcal{R}(E_R)}{F_{A}^2(E_R)}
  \right) \,,
\end{align}
where energies and velocities are related by eq.~\eqref{eq:vm}.  In
agreement with the discussion in sec.~\ref{sec:eta_bound} we see that
the first term on the right-hand side of eq.~\eqref{eq:boundCS_FK}
agrees with eq.~\eqref{eq:RAtb} in the limit of a single target.

In general, eq.~\eqref{eq:boundCS_FK} will lead to a stronger bound on $\rho_\chi
\sigma_{\rm SI/SD}$ than eq.~\eqref{eq:RAtd}. However, it requires a
significantly more precise measurement. The spectrum
$\mathcal{R}(E_R)$ has to be measured with high precision and all
detector effects such as energy resolution and efficiencies have to be
de-convoluted. Certainly this program cannot be carried out in the
case of the 3 events from CDMS-Si, which we used above to illustrate
the bound from eq.~\eqref{eq:RAtd}.  In conclusion, the bound from
eq.~\eqref{eq:boundCS_FK} is useful if a precision measurement of the
DD event spectrum is available, while for low-statistics ``discovery
signals'' the bound from eq.~\eqref{eq:RAtd} can still be applied and
gives a robust lower bound on $\rho_\chi \sigma_{\rm SI/SD}$. Furthermore, the bound of eq.~\eqref{eq:RAtd} can be applied to multi-target detectors, while that of eq.~\eqref{eq:boundCS_FK} cannot, as in general one cannot extract the different $\eta (v_m^A)$ from just one signal. Therefore, in those cases one needs to assume that a particular nuclei gives the dominant contribution.
Let us proceed by comparing the two bounds in the
case of a hypothetical future precision DD measurement.

\subsection{Mock data for a possible DD signal}
\label{sec:mock}

We introduce mock data for a possible future signal in a DD
experiment, also in view of the discussion related to LHC following
below. For SI interactions present limits from DD are so
strong that constraints from LHC are typically not competitive, while for SD
interactions LHC and DD are probing a similar region in parameter
space. Therefore we will concentrate on SD
interactions in this section. To generate mock data for a future DD signal we assume
DM with $m_\chi = 150$~GeV and $\sigma^p_{\rm SD} =\sigma^n_{\rm SD}=
5\cdot 10^{-41}\, \rm cm^2$, which is below the current limits
\cite{Aprile:2013doa, Savage:2015xta, Felizardo:2014awa, Choi:2015ara,
  Aartsen:2012kia} but should be observed in the not-too-far future. For SD interactions we take the nuclear structure functions from ref.~\cite{Bednyakov:2006ux}.  

As a representative example we consider a future xenon based
experiment \cite{Malling:2011va,Baudis:2012bc,Aprile:2012zx}.  We
adopt a threshold of $3$~keV and take natural abundances of the
isotopes with spin $^{129}$Xe ($26.4\,\%$) and $^{131}$Xe
($21.2\,\%$). We neglect the small mass difference between
  the two xenon isotopes, which implies that $v_m$ and hence also
  $\eta(v_m)$ becomes independent of the isotope. We simulate mock
data assuming the SHM (see footnote~\ref{shm}) and a local DM density
$\rho_\chi = 0.4$~GeV/cm$^3$. For an exposure of 1~ton~yr at 100\%
efficiency and an energy resolution of 1~keV approximately $77$ events
would be observed in the energy range $3-45$~keV.  In the following
analysis we compute the $90\%$~CL lower bound in this energy
range. Notice that we neglect a possible contamination with background
and systematic errors. This idealized analysis suffices to illustrate
the power of our bound. Once applied to real data an appropriate
statistical analysis will have to be performed.
 
\begin{figure}
	\centering
	\includegraphics[width=0.72\textwidth]{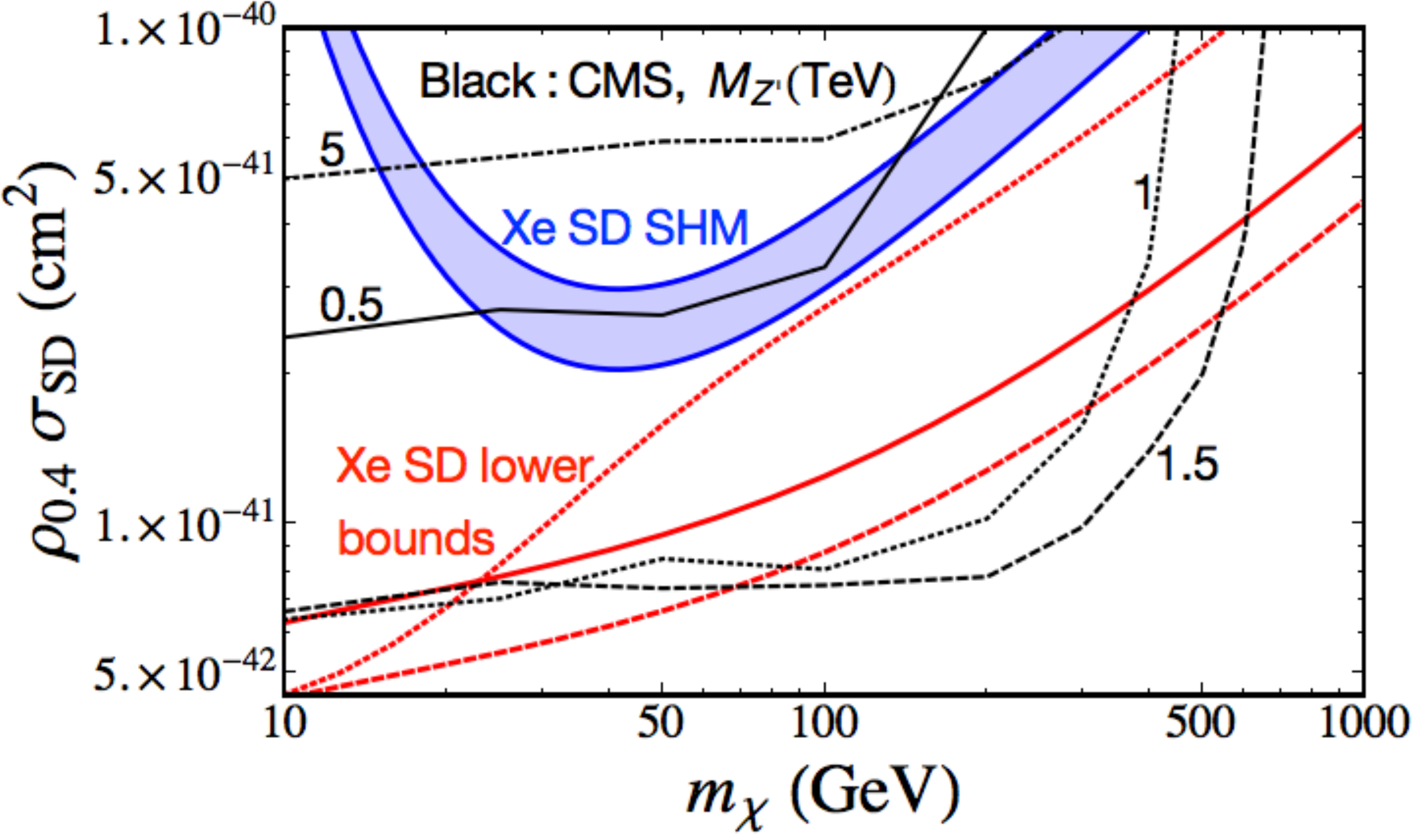}
	\caption{Red curves show the lower bound on $\rho_{0.4} \,
          \sigma_{\rm SD}$ for the mock data generated for a xenon
          experiment, with $\rho_{0.4} \equiv \rho_\chi/(0.4 \, \rm
          GeV\, cm^{-3})$. The red solid, dashed and dotted curves correspond
          to the bounds from eqs.~\eqref{eq:RAtd}, \eqref{eq:RAtb}, and
          \eqref{eq:boundCS_FK}, respectively. The blue-shaded region
          corresponds to the allowed range assuming the SHM. Black
          curves show upper limits from CMS at 95\% CL assuming the simplified
          Majorana DM model for different masses of the $Z'$ mediator
          (labels in the plot give $M_{Z^\prime}$ in TeV).} \label{DDvsLHC}
\end{figure}

In fig.~\ref{DDvsLHC} we show the lower bounds on
$\rho_\chi\sigma_{\rm SD}$ resulting from this assumed DD signal,
based on the bounds from eq.~\eqref{eq:RAtd} (solid red),
eq.~\eqref{eq:RAtb} (dashed red), and eq.~\eqref{eq:boundCS_FK}
(dotted red).  These bounds can be compared to the region obtained
from assuming the SHM (blue-shaded band). This region is obtained by
simply fitting the total number of predicted events in the full energy
range and is therefore a band degenerate in mass. This approach has
been adopted in order to compare to the lower bound based on the same
information. We note that if the SHM is assumed information on
$m_\chi$ can be extracted by performing a spectral fit. In
  comparing the curves one should keep in mind that both the dotted
  and the dashed curves assume a perfect determination of the spectrum
  in an idealized experiment and correspond to the infinite-statistics
  limit. In contrast, the solid red curve and the blue region (SHM)
  show the $90\%$~CL based on the statistical error from the 77
  expected events.

By comparing the dashed and the dotted curves we appreciate the
different strengths of the bounds based on eq.~\eqref{eq:bound} and
\eqref{eq:bound_FK}, respectively. We observe that they merge at low
DM masses, in agreement with fig.~\ref{vmeta}, where, for large
velocities (as relevant for small DM mass), the two bounds come close
to each other and the bounds become weak in all cases. For masses
$m_\chi \gtrsim 50$~GeV the limit from eq.~\eqref{eq:boundCS_FK}
(based on \eqref{eq:bound_FK})~\cite{Feldstein:2014ufa} clearly
becomes stronger than the one from eq.~\eqref{eq:RAtb} (based on
\eqref{eq:bound}), and comes relatively close to the ``true''
region. Again those features follow from the behavior shown in
fig.~\ref{vmeta}. Note that both of those curves (dashed and dotted
red) ignore effects of energy resolution and assume a perfect
measurement of the spectrum $\mathcal{R}(E_R)$ at infinite precision
(for the dashed curve we evaluate the bound of eq.~\eqref{eq:RAtb} at
the threshold of 3~keV).  

The red solid curve corresponds to the bound from eq.~\eqref{eq:RAtd}
based on the total event rate in the full energy range, including also
the finite energy resolution of 1~keV. The energy resolution is also
the reason why this bound is stronger than the ``ideal'' bound from
eq.~\eqref{eq:RAtb} (dashed): because of the energy smearing
events from below the threshold are reconstructed within the analysis
window. This is a well-known effect, in particular in the context of the
sensitivity to low-mass DM, and it turns out also to be important for the
bound discussed here. The reason why the slopes of the
  red dotted curve and the SHM region for $m_\chi \gtrsim 100$~GeV are
  slightly different is also the effect of the finite energy
  resolution. We have checked though, that all bounds as well as the
  SHM region become parallel for $m_\chi \gtrsim 1$~TeV, as expected
  from the $1/m_\chi$ dependence (irrespective of resolutions) in the
  limit of $m_\chi \gg m_A$.

%%%%%%%%%%%%%%%%%%%%%%%%%%%%%%%%%%%%%%%%%%%%%%%%%%%%%%%%%%%%%%%%%%%%%%%%%%%%%%%
\section{Comparison of a direct detection signal with LHC limits}
%%%%%%%%%%%%%%%%%%%%%%%%%%%%%%%%%%%%%%%%%%%%%%%%%%%%%%%%%%%%%%%%%%%%%%%%%%%%%%%
\label{sec:LHC}

The comparison of a signal in a DD experiment with data from a
collider experiment as well as the consideration of the hypothesis of
a thermal history of the DM candidate necessarily depend on the particle physics
model, since different particle reactions are relevant. 
In this section we adopt a specific simplified model for the DM
candidate to illustrate how the halo bounds applied to a possible future DD
signal can be used in the context of limits from LHC. 
In sec.~\ref{sec:relic}, we will use the same simplified model to discuss
a consistency check for the thermal freeze-out hypothesis.

In so-called simplified models for DM, a DM candidate
  particle (assumed to be stable) and a force mediator are added to
  the Standard Model, see refs.~\cite{Abdallah:2014hon, Malik:2014ggr} for
  a summary of the current status.
As an example, we here adopt a simplified model with a
Majorana fermion $\chi$ as the DM candidate, which interacts
with the SM quarks $q$ via a $Z'$ boson with
axial-vector couplings. The interaction Lagrangian of this model is
\begin{align}\label{eq:Lint}
\mathcal{L}_{int} = g_{\chi} \bar{\chi}\gamma_\mu  \gamma^5\chi Z'^{\mu} +  g_{q} \bar{q}\gamma_\mu \gamma^5q Z'^{\mu} \;,
\end{align}
where $g_\chi$ and $g_q$ are the strengths of the $Z'$ interaction
with the dark matter and light quarks, respectively. We assume equal
couplings $g_q$ to $u,d,s,c$ quarks. Couplings to the
  third generation are irrelevant for DD (see below), and have little
  impact on LHC phenomenology (with the exception of on-shell
  production of the $Z'$, where the mono-jet rate depends on the
  partial widths of the $Z'$ \cite{Chala:2015ama}).  This simple framework suffices
  to discuss the phenomenology of interest to us; an extensive
  analysis of the model is beyond the scope of this work. Similar
  models have been considered recently for instance in
  refs.~\cite{Arcadi:2013qia, Lebedev:2014bba, Duerr:2014wra,
    Buchmueller:2014yoa, Chala:2015ama}.

The spin-dependent scattering cross section is given by
\begin{align}\label{eq:sigSD_Zp}
\sigma_{\rm SD}^N=\frac{12}{\pi}\frac{g_\chi^2 }{M_{Z^\prime}^4 } \mu^2_{\chi p} \left(\sum_q g_q\Delta_q^N \right)^2\, ,
\end{align}
where $N$ may denote a proton, $p$, or a neutron, $n$. The spin
coefficients, which parametrize the contribution of the quark species
$q$ to the spin of the nucleon, are given by $\Delta_u^p=\Delta_d^n=
0.84$, $\Delta_d^p= \Delta_u^n= -0.43$ and
$\Delta_s^{p,n}=-0.09$~\cite{Belanger:2008sj}. Note that
  for our choice of equal couplings to quarks there will be a negative
  interference between the up and down quark contributions. Hence the
  scattering cross section is sensitive to the particular choice of
  $g_q$ (including their relative signs).

Both ATLAS and CMS have obtained stringent limits on the interactions
of dark matter with Standard Model (SM) particles based on monojet
searches~\cite{Aad:2015zva,Khachatryan:2014rra}. Here, we derive an
upper limit on $\sigma_{\rm SD}$ from the 95$\%$~CL upper limit on anomalous monojet production
reported by the CMS collaboration~\cite{Khachatryan:2014rra} based on
$19.7\; \mbox{fb}^{-1}$ collected at $\sqrt{s}=8$
TeV.\footnote{Additional constraints on the model coming from
    di-jet searches are discussed in ref.~\cite{Chala:2015ama}.}
MonteCarlo samples of the process $p p \rightarrow \chi\chi + jet$
generated with CalcHEP~\cite{Belyaev:2012qa}, are passed to
Pythia~\cite{Sjostrand:2007gs} for hadronization before we simulate
the effect of the CMS detector with
Delphes~\cite{deFavereau:2013fsa}. As a cross check we have reproduced
the CMS limits for dark matter interacting with quarks via effective
operators. We find that the difference between our results and the official CMS
limits, which can be seen as an estimate of the systematic uncertainty of our reinterpretation, is always smaller than $20\%$.  For our CMS mono-jet analysis we keep the
width of the $Z'$ constant at a typical value for the considered
parameter range.  In general the width can be expected to influence
the LHC limits. However, a full recast of the CMS search is beyond the
scope of this work.  In the fixed width approximation, LHC signatures
depend only on the product of the couplings $g_\chi g_q$.

Note that within our assumption of equal couplings to
light quarks, $g_q$ can be pulled out of the sum in
eq.~\eqref{eq:sigSD_Zp} and $\sigma_{\rm SD}$ depends only on the
product $g_\chi g_q$. Hence, for fixed DM and mediator masses, a DD
signal provides a lower limit on $g_\chi g_q$, while LHC sets an upper
limit on this quantity. 

The black curves in Fig.~\ref{DDvsLHC} show the CMS limits on the
Majorana fermion DM with axial interactions for different masses of
the mediator ($Z'$). Comparing these upper limits with the lower bound
from the assumed DD signal (e.g., red solid curve) we find that the
interpretation of such a DD signal in terms of this model is in
conflict with LHC null results for $M_{Z^\prime} = 1 \sim 1.5$~TeV, whereas a
lighter or a heavier $Z'$ could accommodate both results. 
\begin{figure}
	\centering
	\includegraphics[width=0.6\textwidth]{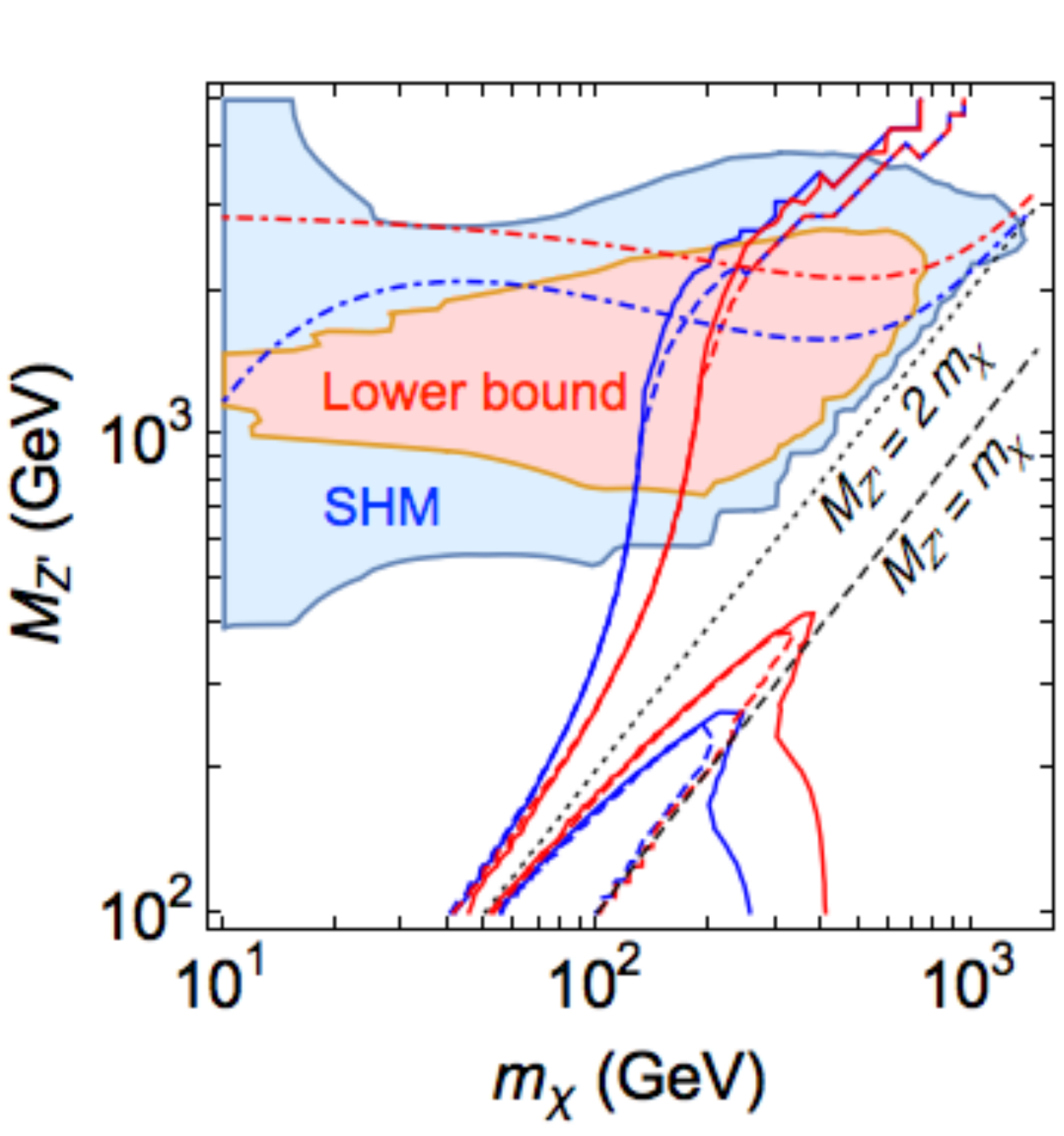}	
	\caption{Constraints in the plane of DM ($m_\chi$) and
          mediator ($M_{Z^\prime}$) masses for the simplified DM
          model, eq.~\eqref{eq:Lint}, assuming a DD signal in a future
          Xe experiment. The colored regions (blue for SHM, red for lower bound) inside the solid curves are
          excluded by comparing the CMS upper limits from mono-jet
          searches to DD data. To the right of
          the solid/dashed curves the DM candidate $\chi$ under
          consideration cannot be a thermal relic, where for the
         solid (dashed) curves we assume $g_\chi = g_q$ ($g_\chi= 10
         \,g_q$). Above the dotted-dashed curves the DD signal can only be achieved if $\Gamma_{\rm Z'}>M_{\rm Z'}/2$. Red curves are based on the bound
          eq.~\eqref{eq:RAtd} and blue ones assume the
          SHM. We take $\rho_\chi = 0.4$~GeV/cm$^3$.} \label{Xe}
\end{figure}
We illustrate this behavior further in fig.~\ref{Xe}, where we confront the interpretation of the  direct detection signal using our velocity independent bound (red) or the SHM (in blue) together with the LHC. As can be seen, the LHC limits exclude a large portion of the parameter space in the $m_\chi-M_{Z^\prime}$ plane independent of the velocity distribution. Furthermore, one has to take into account that in any sensible model the total width of the particles should be significantly smaller than their masses. We illustrate this in fig.~\ref{Xe}, where above the dotted-dashed curves the DD signal implies that $\Gamma_{\rm Z'}>M_{\rm Z'}/2$. In this region the interpretation of the signal in terms of the simplified model is questionable and should be taken with caution.

\begin{figure}
	\centering
  \includegraphics[width=0.65\textwidth]{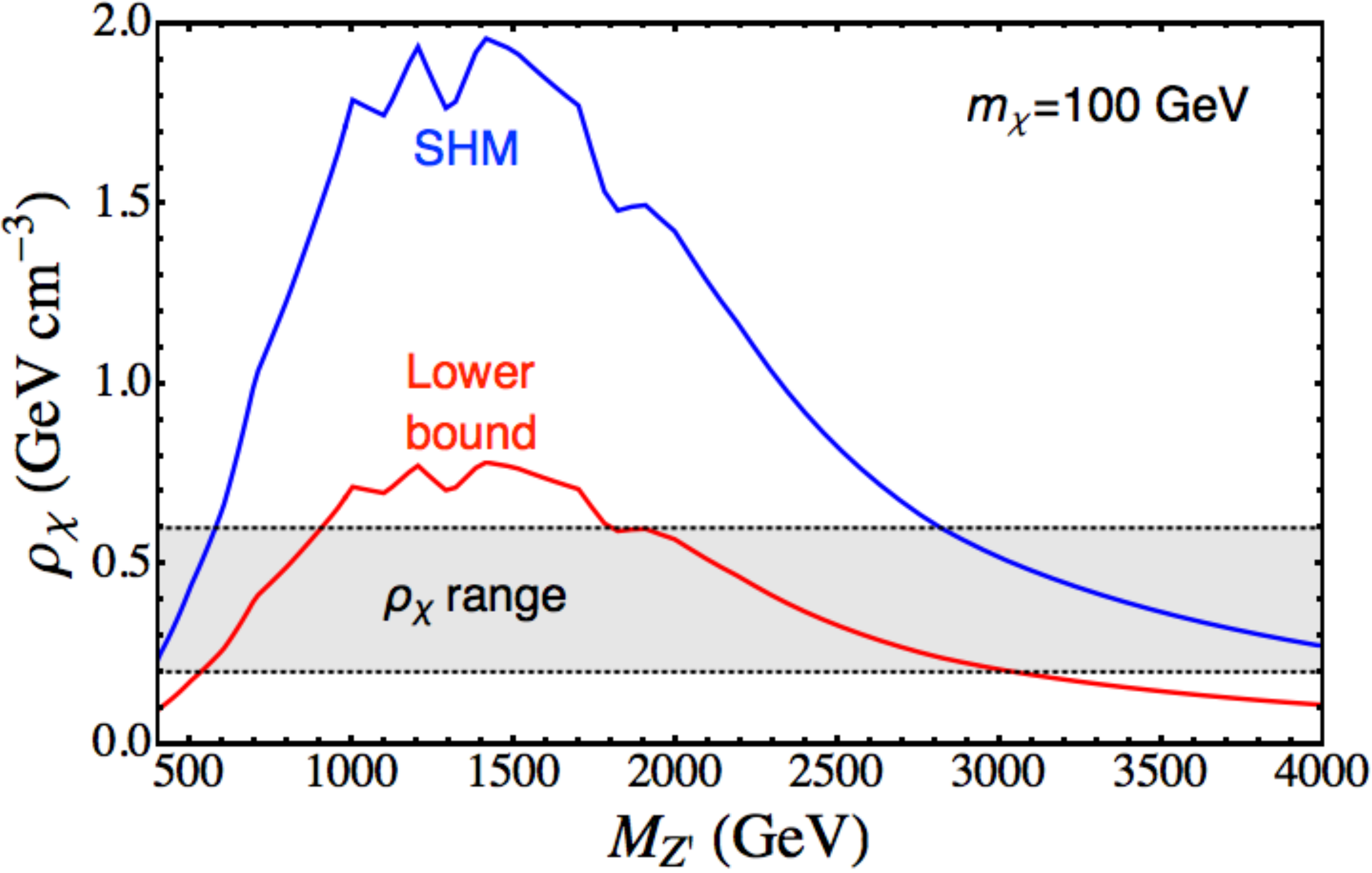}	
  \caption{Lower bound on the local DM density $\rho_\chi$ from the
    combined lower bound on the scattering cross section from the DD
    mock data signal and the upper limits from CMS mono-jet searches
    within the simplified DM model. Lower bounds are shown as a
    function of the mediator mass for a fixed DM mass of 100~GeV. The
    red curve uses the velocity distribution independent lower bound
    from DD, whereas the blue curve assumes the SHM. The grey shaded
    horizontal band indicates the preferred range for $\rho_\chi$ from
    Milky Way observations.} \label{fig:rho_mZ}
\end{figure}

For figs.~\ref{DDvsLHC} and \ref{Xe} we have assumed $\rho_\chi = 0.4
\, \rm GeV \, cm^{-3}$. For different values of the local DM density
the bounds from DD would shift, whereas the LHC limits would remain
unaffected. Hence, the combined lower bound from DD and upper bound
from LHC can be re-cast into a lower bound on the local DM
density. This is shown in fig.~\ref{fig:rho_mZ} for a fixed DM mass of
100~GeV as a function of the $Z'$ mass. The region in parameter space
of the model where these limits are larger than allowed by
astrophysical determinations of $\rho_\chi$ are
excluded.\footnote{The wiggles in the red and blue curves in
  fig.~\ref{fig:rho_mZ} are numerical artefacts related to the Monte
  Carlo statistics of the CMS detector simulation.}
Note that throughout this section we have compared an assumed future
signal from a DD experiment with current LHC limits, while future
limits are expected to increase the sensitivity by up to an order of
magnitude~\cite{ATL-PHYS-PUB-2014-007}.

%%%%%%%%%%%%%%%%%%%%%%%%%%%%%%%%%%%%%%%%%%%%%%%%%%%%%%%%%%%%%%%%%%%%%%%%%%%%%%%
\section{Confronting a direct detection signal with the thermal freeze-out hypothesis}
%%%%%%%%%%%%%%%%%%%%%%%%%%%%%%%%%%%%%%%%%%%%%%%%%%%%%%%%%%%%%%%%%%%%%%%%%%%%%%%
\label{sec:relic}

Under the assumption that a DM candidate $\chi$ has been in thermal
equilibrium with the plasma in the early Universe its relic abundance
will be determined by the freeze-out of the $\chi\chi$ annihilation
processes:\footnote{Notice that in the quantitative analysis at the end of this section we do not use the approximate relation from eq.~\eqref{eq:FO}, but we compute $\Omega_{\chi} h^2$ numerically using \mbox{micrOMEGAs}~\cite{Belanger:2013oya}.}
\begin{equation}\label{eq:FO}
  \Omega_\chi h^2 \approx \Omega_{\rm tot} h^2 \, \frac{\langle \sigma_{\rm th} v \rangle}
   {\langle \sigma_{\chi\chi} v \rangle} \,,
\end{equation}
where $\Omega_\chi$ is the abundance of $\chi$ relative to the critical
density of the Universe today, $h$ parametrizes the Hubble
constant, we use $\langle \sigma_{\chi\chi} v \rangle$ to denote the total
annihilation cross section of $\chi$ times velocity, averaged over the
thermal distribution in the early Universe, and $\langle \sigma_{\rm
  th} v \rangle \approx 3 \cdot 10^{-26}\rm \,cm^3\,s^{-1}$ is the annihilation cross
section required to obtain the DM abundance as determined
from cosmological observations~\cite{Ade:2015xua}, $\Omega_{\rm tot} h^2 = 0.1194 \pm
0.0022$.

In a given particle physics model, the scattering cross section can be
related to the annihilation cross section. Hence, under the thermal
freeze-out hypothesis, a lower bound on $\rho_\chi \sigma_{\rm SI/SD}$ will
provide a lower bound on $ {\langle \sigma_{\chi\chi} v \rangle}$ and therefore an upper bound on the relic density via eq.~\eqref{eq:FO}. For a given DM halo model this upper bound on the relic density becomes an equality. If this upper bound on the energy density is smaller than the value
for $\Omega_{\rm tot}$ determined from cosmological
observations (or equivalently if the lower bound on $\langle \sigma_{\chi\chi} v \rangle$ is larger than $\langle \sigma_{\rm th} v \rangle$), the observed direct detection rate
 is inconsistent with the thermal production of the DM candidate within a given model. 
  
One might wonder whether it is possible to avoid this conclusion by
allowing $\chi$ to become a subdominant component of DM. However, we
would expect naively that a cosmological subdominant component of dark
matter with $\Omega_\chi < \Omega_{\rm tot}$ does not constitute all
the dark matter locally and therefore $\rho_\chi < \rho_{\rm tot}$. While a completely general
statement is not possible we find that our bound can be extended to
this case under certain conditions: 
\begin{enumerate}
\item Only a single subdominant
species $\chi$ induces the DD signal while more particles contribute
to the DM in the universe (for instance well-known examples are axions or keV-scale sterile
neutrinos).  
\item The local density of $\chi$ in the galaxy is proportional to the global density:
  \begin{equation}\label{eq:density}
    \frac{\rho_\chi}{\rho_{\rm tot}} = 
    \frac{\Omega_\chi h^2}{\Omega_{\rm tot} h^2} \,.
  \end{equation}
 \end{enumerate}
Eq.~\eqref{eq:density} assumes that all DM components contributing to structure formation are cold. In the presence of a cold/warm DM mix this assumption may be violated, see ref.~\cite{Anderhalden:2012qt} for a
  numerical study. Implications for the DD and LHC comparison under
  the proportional assumption of eq.~\eqref{eq:density} have been
  discussed previously in ref.~\cite{Bertone:2010rv}.

Typically we expect $\langle \sigma_{\chi\chi} v \rangle \propto \sigma_{\rm SI/SD}$, and, as the lower bound on $\sigma_{\rm SI/SD}$ scales as $1/\rho_\chi$, see eq.~\eqref{eq:RAtd}, the upper bound on $\Omega_\chi$ will be proportional to $\rho_\chi$. Thus, if the upper bound on $\Omega_\chi$ is smaller than $\Omega_{\rm tot}$ for $\rho_\chi =
\rho_{\rm tot}$, eq.~\eqref{eq:density} implies that it will also be violated for any other value of
$\Omega_\chi < \Omega_{\rm tot}$ and $\rho_\chi<\rho_{\rm tot}$. Hence, under these assumptions $\chi$ is inconsistent with having a thermal abundance, irrespective of whether it provides all of the DM or only part of it. This argument can be avoided by invoking some exotic physics which breaks the scaling relation in eq.~\eqref{eq:density} and enhances the local density of $\chi$ relative to the other DM species. The naive relic density approximation used in this discussion can be avoided by combining the lower bound
of eq.~\eqref{eq:RAtd} on $\rho_\chi\,\sigma_{\rm SI/SD}$ directly with eq.~\eqref{eq:density}. This yields a lower bound on $ \Omega_\chi \,\sigma_{\rm SI/SD}$ which is completely general and can be used within any given model even if $\Omega_\chi$, $\langle \sigma_{\chi \chi}v \rangle$ and $\sigma_{\rm SI/SD}$ are not related by simple scaling relations or if higher precision is desired.

To illustrate the relic density bound numerically we adopt the DD mock
data from sec.~\ref{sec:mock} and the $Z'$ model from
sec.~\ref{sec:LHC}.  For calculating the relic density we use only the
most minimal model able to provide a relevant scattering cross
section, e.g., taking into account only $Z'$ couplings to the light
quarks (see sec.~\ref{sec:LHC}). If we allow for the possibility of
additional annihilation channels (for instance into third generation
quarks, leptons, or into hidden sector particles beyond the simplified
model) the relic abundance can only become
smaller.\footnote{This statement may not hold close to the
    resonance region, where additional channels lead to a larger width,
    implying a smaller resonant annihilation cross section.} Hence,
using the minimal model to calculate the upper bound on the relic
abundance is conservative, as additional channels will make the
inequality worse.

To the right of
the solid or dashed curves in fig.~\ref{Xe}, $\Omega_\chi = \Omega_{\rm tot}$ is excluded
where the red curve uses the bound from eq.~\eqref{eq:RAtd} and the
blue one assumes the SHM.  In large part of the parameter space the
bound is independent of the relative size of the coupling constants
$g_\chi$ and $g_q$, since the relevant cross sections depend only on
the product $g_\chi g_q$ (see below). To demonstrate this behavior
explicitly we show the bound for $g_\chi = g_q$ (solid) and $g_\chi=
10\, g_q$ (dashed).

For DM masses below the threshold for $Z'$ pair production, i.e. for
$M_{Z^\prime} > m_\chi$, the annihilation cross section scales approximately
as $\sigma_{\chi\chi} v \propto g_q^2 g_\chi^2 m_\chi^2/(M_{Z^\prime}^2- 4
m_\chi^2)^2$, while the scattering cross section from
eq.~\eqref{eq:sigSD_Zp} behaves as $\sigma_{\rm SD} \propto g_q^2
g_\chi^2 m_p^2/M_{Z^\prime}^4$ (approximately independent of $m_\chi$
for $m_\chi \gg m_p$). In this regime both, $\sigma_{\chi\chi} v$ and
$\sigma_{\rm SD}$ depend only on the product $g_\chi g_q$ and not on
$g_\chi$ or $g_q$ individually. This is apparent in the fig.~\ref{Xe},
where for $M_{Z^\prime} > 2m_\chi$ the curves for $g_\chi = g_q$ and
$g_\chi= 10 \,g_q$ essentially overlap. Furthermore, since the bound 
scales approximately as $\sigma_{\rm SD} / \langle \sigma_{\chi\chi}v\rangle$, it follows that
it is independent of $g_\chi g_q$. 

Near the resonance, $M_{Z^\prime} \approx 2m_\chi$, the
annihilation cross sections will be strongly enhanced for a given
scattering cross section, and therefore for a given scattering
cross section the relic density bound becomes very constraining. The structures along the line $M_{Z^\prime}
\approx m_\chi$ in fig.~\ref{Xe} can be understood from the appearance
of the $\chi\chi\to Z' Z'$ annihilation channel in that region which
lead to a different dependence of $\sigma_{\chi\chi} v$ on $g_\chi$
and $g_q$.  As can be seen in fig.~\ref{Xe} the results for $g_\chi =
g_q$ and $g_\chi= 10 \, g_q$ differ significantly in this region.

Finally, we have investigated the impact of a subdominant dark matter species. As expected, the precise value of $\Omega_\chi$ generically has only a
minor impact on the bound. Numerically, the bound changes by less than $20\%$ for $\Omega_\chi/\Omega_{\rm tot} > 0.1$ and by less than a factor of two as long as
 $\Omega_\chi/\Omega_{\rm tot} > 0.01$. 
An even smaller relic density can typically only be achieved if the coupling constants $g_{\chi,q}$, and consequently $\Gamma_{Z'}$, are large. The relation between $\langle\sigma_{\chi\chi} v \rangle$ and $\sigma_{\rm SD}$ is more complicated in this case and $\Omega_\chi \,\sigma_{\rm SD}$ exhibits a non-trivial scaling behavior. 

%%%%%%%%%%%%%%%%%%%%%%%%%%%%%%%%%%%%%%%%%%%
\section{Discussion and conclusions} 
%%%%%%%%%%%%%%%%%%%%%%%%%%%%%%%%%%%%%%%%%%%
\label{sec:conclusions}

We have derived lower bounds on the product of the DM--nucleus
scattering cross section and the local DM density from a positive
signal in a direct detection experiment, which is independent of the
DM velocity distribution. If an upper bound on the local DM density
from kinematical Milky Way observations is applied, our bounds provide
a robust lower bound on the scattering cross section.

We have discussed different versions of such bounds. One of them is
based only on the number of events observed in a certain recoil energy
interval and leads to a robust bound even in the case of few signal
events. As illustration we have applied this bound in the context of
the 3 candidate events found in CDMS silicon data.  A second version
requires an accurate measurement of the recoil spectrum, including a
deconvolution of resolution and efficiency factors, however, it provides
more stringent lower bounds on the cross section.

In this work we have restricted the analysis to
  time-averaged signals in direct detection experiments, neglecting
  the small annual modulation effect. In ref.~\cite{Herrero-Garcia:2015kga} it is
  shown that also the annual modulation signal can be used to obtain a
  halo-independent lower bound on the scattering cross section, in
  particular in combination with the methods developed in
  refs.~\cite{HerreroGarcia:2011aa, HerreroGarcia:2012fu}.

In order to illustrate our bounds we have assumed the observation of a signal in just one direct detection experiment. Let us briefly comment on the case of a positive signal in
  more than one experiment, using different target nuclei. A priori
  our bound can be calculated for each experiment and it may happen
  that depending on the DM mass different experiments provide the
  strongest bound. However, in the lucky case of a multiple DM
  detection, more information is available and other methods may be
  more appropriate. First, one may try to answer the question of
  whether the signals are consistent with each other in a
  halo-independent way \cite{Drees:2007hr, Drees:2008bv, Fox:2010bz,
    Fox:2010bu, McCabe:2011sr, McCabe:2010zh, Frandsen:2011gi,
    HerreroGarcia:2011aa, HerreroGarcia:2012fu, DelNobile:2013cta,
    DelNobile:2013cva, Bozorgnia:2013hsa, Cherry:2014wia, Fox:2014kua,
    Feldstein:2014gza, Feldstein:2014ufa, Bozorgnia:2014gsa,
    Anderson:2015xaa,Scopel:2014kba,Kavanagh:2012nr,Kavanagh:2013wba}. Assuming
  that they are consistent, the methods of ref.~\cite{Drees:2008bv}
  can be used to extract the DM mass in a halo-independent way, see
  also \cite{Kavanagh:2013wba}. This DM mass can then be used for the
  bounds on the cross section discussed here. Finally, more
  information on particle physics can be obtained. For instance one
  can try to infer the relative coupling strength to neutrons and
  protons from the data simultaneously to bounding the cross
  section. A detailed investigation of the multi-detection case is
  beyond the scope of this work.
  
For a given particle physics model, a lower bound on the scattering
cross section from direct detection data can be compared to data from
LHC. As an example we consider a so-called simplified DM model and
derive allowed regions in the model parameter space by the comparison
of an assumed signal in a future direct detection experiment with
upper limits from LHC mono-jet searches.

Finally we have shown how our bounds from a direct detection signal
can be used to test the hypothesis that the particle responsible for
the signal is a thermal relic. Furthermore, the bound can be used to formulate a condition which
has to be fulfilled under the assumption of a thermal history of the
DM candidate, irrespective of whether this particle provides all of
the DM or only part of it. Again we have used a simplified DM model as
an example, and have identified the region in the space of DM and
mediator masses, which would exclude the thermal freeze-out mechanism
for an assumed direct detection signal.

While in this work we have used a simple DM model consisting of a
Majorana fermion as DM interacting with the Standard Model via a
$Z'$ mediator, we note that our bounds can be applied for any other
model which allows to relate the scattering cross section to LHC
observables and the relic abundance. In our example model all
observables depend only on four parameters (DM and mediator masses and
two couplings), in large part of the parameter space only on three
(only the product of the two couplings is relevant). In more
complicated models with more parameters a marginalization over some
parameters (or optimization of the inequalities) will have to be
performed.

In the same way we used our halo bounds for the comparison of a direct
detection signal with LHC limits, it is also possible to confront a
direct detection signal with limits from indirect detection (searching
for DM annihilation products from astrophysical environments like
dwarf galaxies or the galactic centre). For the specific $Z'$ model
used as an example in this work, the annihilation cross section
$\langle \sigma v \rangle$ is dominated by $p$-wave processes and,
consequently, the annihilation rate today is strongly
suppressed. Therefore, we expect limits from indirect detection to be
very weak for this model.

To conclude, we want to encourage the community to show the DM direct
detection positive results, which hopefully will occur at some point
in the near future, using the velocity distribution independent lower
bound on the cross section derived here, in addition to the usually
assumed Maxwellian halo model.

\bigskip

{\bf Note added:} After the completion of this work and
  submission to the arXiv, the preprint ref.~\cite{Ferrer:2015bta}
  appeared, where also a halo-independent lower bound on the
  scattering cross section is derived.

\bigskip

{\bf Acknowledgements:} We would like to thank Felix Kahlhoefer for
helpful discussions. The work of MB was partially supported by the
G\"oran Gustafsson Foundation. TS and SV acknowledge support from the
European Union FP7 ITN INVISIBLES (Marie Curie Actions,
PITN-GA-2011-289442).

\bibliographystyle{JHEP.bst}
\bibliography{dir-bound}

\end{document}